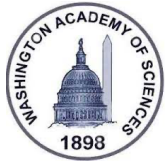

# THE SEARCH FOR EXTRASOLAR MOONS:
## PHOTON FLUX PERTURBATIONS OF KEPLER TRANSIT LIGHT CURVES

ANTONIO PARIS
PLANETARY SCIENCES, INC.

**ABSTRACT**

In this study, we expand the discussion of extrasolar planetary research by proposing a new approach to detecting extrasolar moons using Kepler (K2) light curves. We shaped this investigation by comparing transit light-curve data of $5 \times 10^3$ main-sequence stars cataloged in NASA's Exoplanet Archive, Radio Galaxy Zoo's Exoplanet Explorers and the Exoplanet Follow-up Observing Program (ExoFOP), which served as the repository to collect and analyze supplementary K2 data. By examining K2 light curves, various characteristics related to transits were modeled, which we then compared with confirmed extrasolar planets, variable stars such as eclipsing binaries and noise or gaps in the data. For illustration, perturbations in the timing of two separate transits for 2MASS J08251369+1425306 ($R_S \approx 0.346_\odot$) produced two characteristic decreases in the photon flux ($\Delta F$) followed by two increases, inferring the presence of an extrasolar planet ($R_P \approx 0.0129_\oplus$) and its companion, an extrasolar moon ($R_M \approx 0.0048_\oplus$). To test our hypothesis, we scrutinized various competing assumptions to resolve the source of the duo-photon flux, namely the *Rossiter–McLaughlin* effect, limb darkening, a multiplanet system or a planet occulting one or more sunspots on the surface of the star. However, we uncovered no analogous light curve in K2 to fit the competing hypotheses and account for the proposed extrasolar planet's companion.

## INTRODUCTION

Launched in 2009, the Kepler spacecraft was designed to survey our region of the Milky Way galaxy to discover Earth-size extrasolar planets in or near the habitable zone and to explore the structure and diversity of extrasolar planetary systems.[1] Kepler, which was launched into an Earth-trailing orbit, monitored the brightness of approximately 530,506 stars in a fixed field of view.[2] Specifically, the spacecraft observed a 100 sq. degree patch of the sky near Cygnus, Lyra and Draco; and rotated by 90 degrees every 90 days to keep the solar panels pointing at the sun.[3] This is also the direction of the Solar System's motion around the center of the galaxy. Thus, the stars that Kepler observed are roughly the same distance from the galactic center as the Solar System, and close to the galactic plane. This fact is important if the position in the galaxy is related to habitability, as suggested by the Rare Earth hypothesis. Kepler's data, moreover, are divided into 90-day quarters.[4] These data were transmitted to Earth, then analyzed to detect periodic dimming (light curves) caused by extrasolar planets when they cross (transit) in front of their parent star. On October 30, 2018, the Kepler spacecraft's end of mission was declared, with the spacecraft finally running out of fuel.[5] During its nine years in orbit, Kepler observed a total of 150,000 main-sequence stars and discovered 2,662 planets, many of which could be promising places for life.[6]

## THE TRANSIT METHOD FOR DETECTING EXTRASOLAR PLANETS

Through the use of photometry, the Kepler spacecraft used the transit method to detect extrasolar planetary systems. This technique consists of regularly observing a star to detect the periodic decrease in luminosity ($\Delta F$), also known as flux, associated with the transit of an extrasolar planet.[7] Both the size of the parent star and the extrasolar planet will determine $\Delta F$ during the transit. For illustration, for a star the size of the Sun, the transit of a Jupiter-sized planet will cause a decrease in the apparent luminosity of ≈1%, whereas this decrease will be ≈0.001% for a planet the size of the Earth.[8] When extrasolar planets pass in front of their host star (as seen from Earth), moreover, a decrease in the photon flux can be measured over time to allow for the construction of a light curve (Figure 1). By measuring $\Delta F$ and knowing the size of the parent star, we can determine the size or radius of the extrasolar planet, its orbital period and, using Kepler's Third Law of Planetary Motion, the average distance of the extrasolar planet from its parent star can be determined.[9]

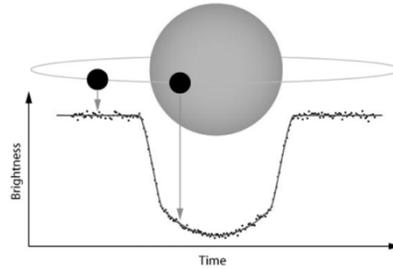

Figure 1: A simulation of a transit light curve as the extrasolar planet blocks light from its parent star. (Source: NASA)

## DATA COLLECTION

**Kepler Spacecraft:** Kepler's primary high reflectance mirror is 1.4 meters (4.6 ft) in diameter with a 100 deg$^2$ (≈12-degree diameter) field of view. The spacecraft's telescope has a Schmidt camera with a 0.95 meter (37.4 in) front corrector plate (lens) feeding a 1.4 meter (55 in) primary mirror. The mirror was specifically designed with sufficient sensitivity to detect relatively small extrasolar planets as they transit their parent star.[10] At the time, the mission objective was a combined differential photometric precision of 20 parts per million on a magnitude 12 star for a 6.5 hour integration.[11] The photometer, moreover, has a soft focus to provide photometry, rather than sharp images. The focal plane of the spacecraft's camera, furthermore, is made from 42 50 mm × 25 mm (2 in × 1 in) CCDs at 2200 × 1024 pixels each, possessing a total resolution of 94.6 megapixels.[12] The data from these pixels were then requantized, compressed and stored in the onboard 16 gigabyte solid-state recorder.

**Kepler 2 (K2) Data Sets:** There are two archives for official Kepler and K2 data products—the Exoplanet Archive, which is hosted at the NASA Exoplanet Science Institute and the Mikulski Archive for Space Telescopes (MAST), which is hosted at the Space Telescope Science Institute (STScI). The Exoplanet Archive primarily hosts data related to the Kepler and K2 mission exoplanet searches.[13] MAST, on the other hand, is responsible for hosting time series data and spacecraft calibration products for Kepler and K2.[14] In addition, this research has made use of the Exoplanet Follow-up Observation Program (ExoFOP), which is operated by the California Institute of Technology, under contract with NASA under the Exoplanet Exploration Program. For this investigation, ExoFOP served as the repository to collect and analyze community-gathered data related to K2, including the light curves presented hereafter.

**2MASS:** The summary of stellar information used in this investigation was obtained using data products from the Two Micron All Sky Survey (2MASS), which is a joint project of the University of Massachusetts and the Infrared Processing and Analysis Center/California Institute of Technology, funded by NASA. The data were collected by 1.3 m telescopes at Mt. Hopkins and CTIO, Chile, which, thereafter, created the Point Source Catalog consisting of over 500 million stars and galaxies, the Extended Source Catalog consisting of 1.6 million resolved galaxies and the Large Galaxy Atlas consisting of ≈ 600 nearby galaxies and globular clusters.[15] The stellar information throughout this study includes galactic coordinates, star magnitude and classification, and distance in parsecs.

**Radio Galaxy Zoo:** Additional light-curve data for this investigation were obtained from Exoplanet Explorers. The archival data are part of Radio Galaxy Zoo—an Internet-crowdsourced citizen science project. Through the use of K2 data, citizen scientists assisted scientists in identifying potential transiting extrasolar planets. The program is hosted by the web portal Zooniverse. To date, data from Kepler's survey revealed over 4,000 candidate extrasolar planets—many of which were identified by citizen scientists participating in the Exoplanet Explorers project.[16]

## COMPARING KEPLER TRANSIT LIGHT CURVES

Numerous astrophysical processes can impersonate planetary transits. Light-curve data sets cataloged in K2 primarily consist of confirmed and/or candidate extrasolar planets, variable stars such as eclipsing binaries and pulsating stars, and noisy data with light curves that have such scattered data points that it is difficult to determine if a valid signal was collected by Kepler. By comparing these light curves, it is then possible to compare the processes at work within each stellar object and/or event.[17] When we plot a new light curve obtained by Kepler, we can compare it to those standard light curves previously observed by other astronomical surveys. Through these methods, therefore, we can attempt to identify the type of object observed by Kepler. For illustration, the light curve for the confirmed extrasolar planet K2-14 b (Figure 2) produced a *flat* light curve with one clear "U-shaped" dip centered at time 0.0.[18] Moreover, when K2-14 b passed behind its parent star at time ≈ 0.13, the light from the extrasolar planet, such as starlight that was reflecting off the planet's surface, was blocked by the parent star. This decrease in brightness is known as a "secondary eclipse."[19]

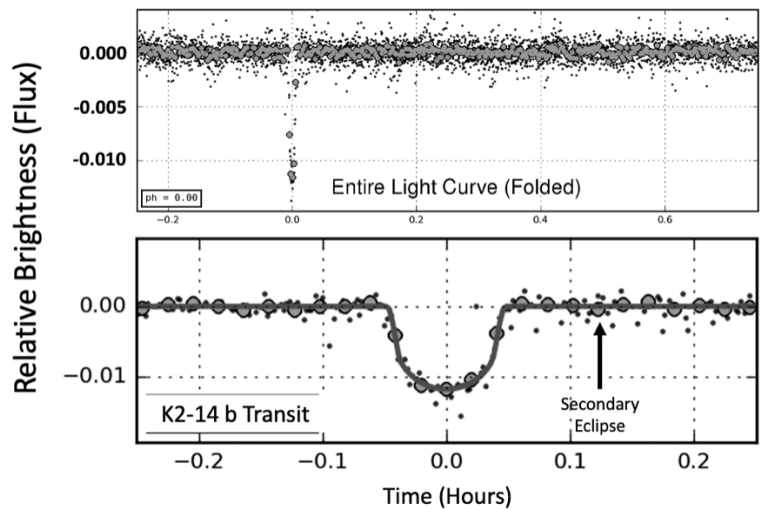

Figure 2: Light curve for K2-14 b
(Source: Exoplanet Explorers and ExoFOP)

Although multiplanet light curves produce comparable "U-shaped" dips in their folded light curve, these systems can be differentiated from single-planet systems by *unfolding* their individual transits. The main-sequence star K2-14, for example, is known to host only one confirmed planet (K2-14 b). When we unfold the light curve for K2-14 b, the data reveal nine separate transits with comparable U-shaped dips at ≈0.0.[20] In contrast, the unfolded light curve for a multiplanet system, such as K2-138 (six known planets), produced transits with dissimilar light curves.[21]

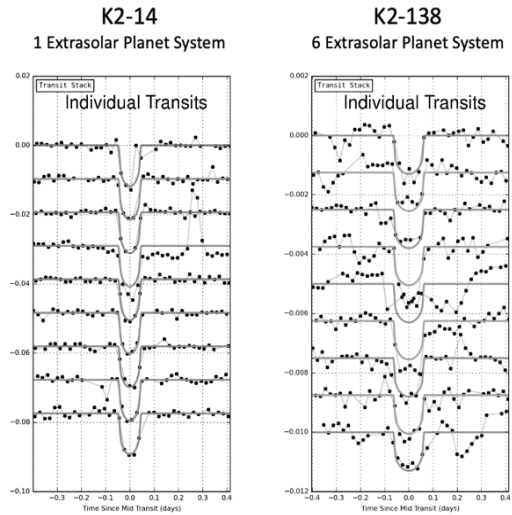

Figure 3: Comparing individual transits for K2-14 and K2-138
(Source: Exoplanet Explorers and ExoFOP)

Some transits are caused by other stars rather than a planet. One type of variable star, such as an eclipsing binary, varies dramatically in brightness in a regular cycle. Eclipsing binaries are a pair of stars that orbit each other—one star will transit in front and behind the other, as seen by Kepler. Their light curves, consequently, will generate *deep dips* and the entire folded light curve will produce a sinusoidal pattern as opposed to a flat light curve. To demonstrate, the light curve for the confirmed eclipsing binary star 2MASS J12232783-0649416 (Figure 4) has a deep minimum when eclipsed at time ≈ 0.0 and a shallower minimum when the dimmer star is eclipsed at ≈0.5 (folded light curve).[22] Furthermore, the folded light curve is sinusoidal as opposed to flat.

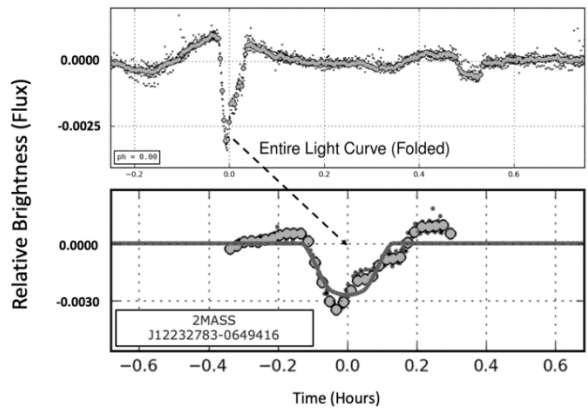

Figure 4: Light curve for 2MASS J12232783-0649416
(Source: Exoplanet Explorers and ExoFOP)

Some stars, such as pulsating variable stars, vary dramatically in brightness in their regular cycle. The radius of this type of star alternately expands and contracts as part of its natural evolutionary aging processes.[23] Their light curves, therefore, are distinctive and show a rapid rise in brightness followed by a smoother decline, shaped like a shark fin.[24] The light curve for the star 2MASS J22515346-1257133 (Figure 5), for instance, shows the characteristic rapid rise at time ≈ 0.2 in brightness when the star expands followed by a more gradual decrease as the star's radius contracts.[25]

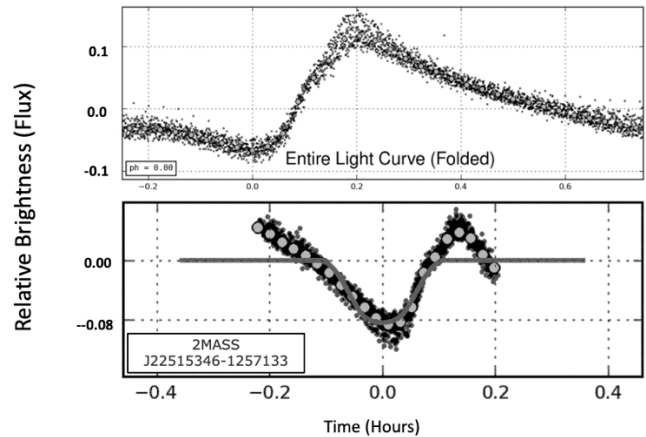

Figure 5: Light curve for 2MASS J22515346-1257133
(Source: Exoplanet Explorers and ExoFOP)

Some light curves have scattered (noisy) data or missing data points so that it is difficult to determine if a valid signal and/or observation was collected by Kepler. The principal source of noise in K2 data was created by the motion of the Kepler spacecraft due to periodic thruster firings. This caused stars to drift across different pixels on the detector, which have varied sensitivity.[26] The light curve for the star 2MASS J03355065+1654303 (Figure 6) visibly displays noise and missing data, and as such, is not an acceptable candidate for extrasolar planet surveys.[27]

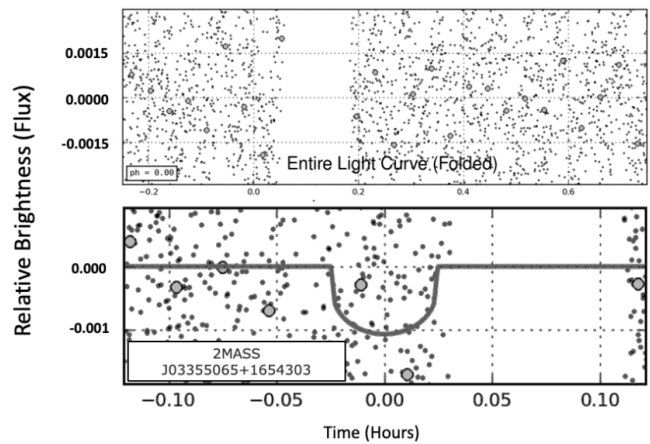

Figure 6: Light curve for 2MASS J03355065+1654303
(Source: Exoplanet Explorers and ExoFOP)

## PROPOSED METHODOLOGY FOR IDENTIFYING EXTRASOLAR MOONS

Extrasolar moons are detectable for the same reasons their parent planets are—they have mass and occupy space.[28] By analyzing K2 perturbations in the timing of transits, it is possible to infer the presence of additional planetary companions, including extrasolar moons.[29] However, before searching for extrasolar moons using K2 data, we first have to consider an essential question: would it be possible to detect an extrasolar moon around a main-sequence star, given the current quality of K2 data? As we demonstrate in our simulation below (Figure 7), we posit, when an extrasolar moon begins to block the light from the parent star (Event A) the initial transit should produce a slight dip in the light curve ($D_A$). Afterward, when the extrasolar planet blocks the starlight during its transit (Event B), it should then produce the typical U-shaped light curve ($D_B$). Similarly, when the extrasolar moon begins to egress from the transit (Event C), an increase in luminosity from the parent star will produce a slight increase in the light curve ($I_C$) followed by a substantial increase ($I_D$) as the extrasolar planet exits the transit (Event D).

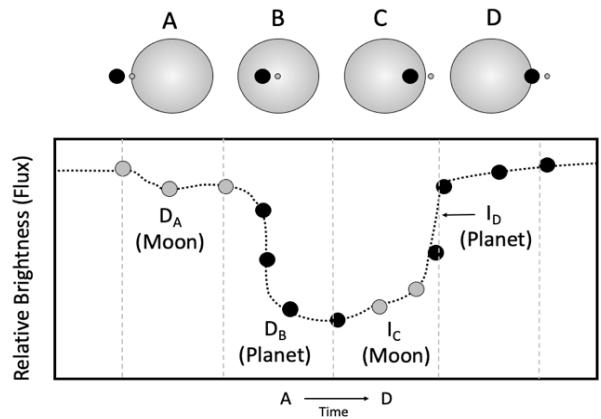

Figure 7: A simulation of transit light curve for an extrasolar planet with a companion moon.
(Source: Planetary Sciences, Inc.)

Assuming the star's radius ($R_S$) is known, the light curve will also provide an *estimate* for the radius of the extrasolar planet ($R_P$) and its moon ($R_M$) by determining the fraction of starlight, $\Delta F$, that was blocked. Although the equation assumes the stellar disc has a uniform brightness, as a preliminary estimate the mathematical relationship (Equation 1) works quite well.[30]

Equation 1

Radius of Planet

$$\Delta F = \frac{R^2_P}{R^2_S} \quad \text{whereas} \quad R_P = R_S \sqrt{\Delta F}$$

Radius of Exomoon

$$\Delta F = \frac{R^2_M}{R^2_S} \quad \text{whereas} \quad R_M = R_S \sqrt{\Delta F}$$

## DATA ANALYSIS AND INTERPRETATION

A random sampling of $5 \times 10^3$ K2 light curves archived in the Exoplanet Explorers project, then cross-referenced with ExoFOP, revealed eight light curves (Table 1) analogous to the extrasolar moon-like perturbations described in our simulation (Figure 7).

| EE Subject ID | EPIC ID (ExoFOP) | 2MASS ID | Ra (deg) | Dec (deg) | Distance (pc) | Radius ($R\_Sun$) | Mass ($M\_Sun$) |
|---|---|---|---|---|---|---|---|
| 7775826 | 211591558 | J08251369+1425306 | 126.307076 | 14.42519 | 312.9 | 0.346 | 0.356 |
| 7608624 | 201456770 | J12224737-0007187 | 185.697449 | -0.121909 | 5.82E+02 | 0.926 | 0.891 |
| 7654197 | 210924845 | J04042735+2147014 | 61.113998 | 21.78367 | 108.4 | 1.549 | 1.192 |
| 7774537 | 211430475 | J08512348+1206040 | 132.847855 | 12.101128 | 5.62E+02 | 0.308 | 0.329 |
| 7775208 | 211527627 | J08142873+1331468 | 23.619728 | 13.529658 | 179.3 | 0.446 | 0.503 |
| 7775402 | 211545599 | J08170774+1346494 | 124.282227 | 13.780388 | 2289 | 28.395 | 1.124 |
| 7605568 | 201192009 | J12081933-0417229 | 182.080521 | -4.289704 | 3.16E+03 | 4.955 | 0.893 |
| 7775832 | 211592036 | J08253940+1425538 | 126.414192 | 14.431643 | 641 | 1.456 | 1.168 |

Table 1: List of eight light curves exhibiting extrasolar moon-like perturbations.
(Source: Exoplanet Explorers (EE Subject ID) and ExoFOP (EPIC ID).

Consider the light curve for 2MASS J08251369+1425306 (K2 Campaign: 5), a main-sequence star located ≈319 $pc$ from Earth (Figure 8).[31] As the proposed extrasolar moon began to block the light from the star (Event A) at time –0.1 the transit generated a minor dip ($\Delta F \approx 0.0002$) in the light curve. Subsequently, when the proposed extrasolar planet blocked the starlight during its transit (Event B), it then produced the corresponding U-shaped light curve at phase 0.0 ($\Delta F \approx 0.0014$). Afterward, when the proposed extrasolar moon exited its transit of the parent star (Event C), the increase in luminosity generated a minor increase in the light curve, followed by an increase in the flux when the extrasolar planet exited the transit (Event D).

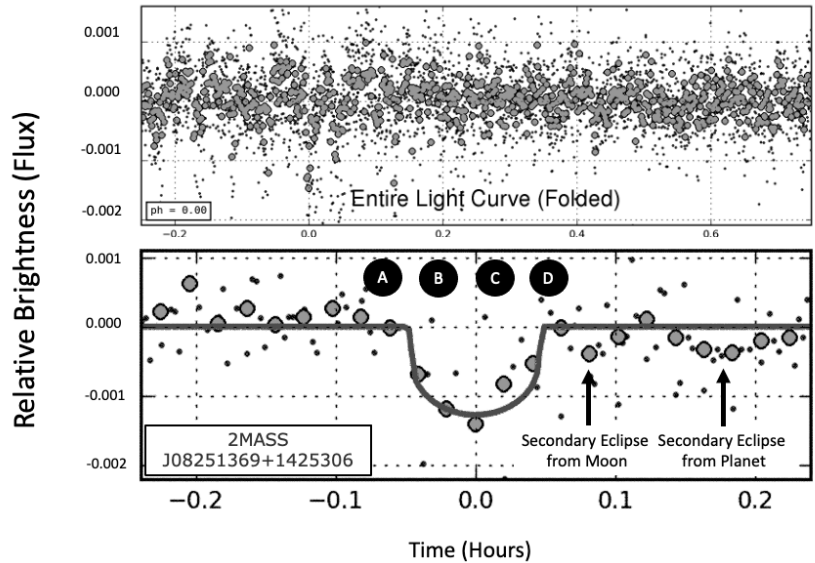

Figure 8: Light curve data for 2MASS J08251369+1425306 with corresponding events A–D.
(Source: Exoplanet Explorers and ExoFOP)

With reference to Equation 1, because the radius for 2MASS J08251369+1425306 is known (Table 1), we can estimate the radius of the extrasolar planet (Equation 2) using the fraction of starlight, $\Delta F \approx 0.0014$, that was blocked (Figure 8) at transit times –0.06 to 0.0.

Equation 2

$$R_P = R_S \sqrt{\Delta F} \quad \text{whereas} \quad R_P = .346 \sqrt{.0014}$$

$$R_P \approx .0129$$

Correspondingly, we can also estimate the radius of the extrasolar moon (Equation 3) using the fraction of starlight, $\Delta F \approx 0.0002$, that was blocked (Figure 8) at transit times –0.1 to –0.06.

Equation 3

$$R_M = R_S \sqrt{\Delta F} \quad \text{whereas} \quad R_M = .346 \sqrt{.0002}$$

$$R_M \approx .0048$$

The folded light curve, furthermore, exhibits a pair of secondary eclipses that we infer are associated with when the extrasolar moon and planet were occulted by 2MASS J08251369+1425306. The first and shorter secondary eclipse at time ≈ 0.08 (Event E) is the light from the extrasolar moon's dayside being blocked as it makes its orbit *behind* the star (Figure 9). Similarly, a subsequent and lengthier secondary eclipse was generated at time ≈ 0.17 (Event F) when the extrasolar planet (which is larger) made its orbit *behind* the star.

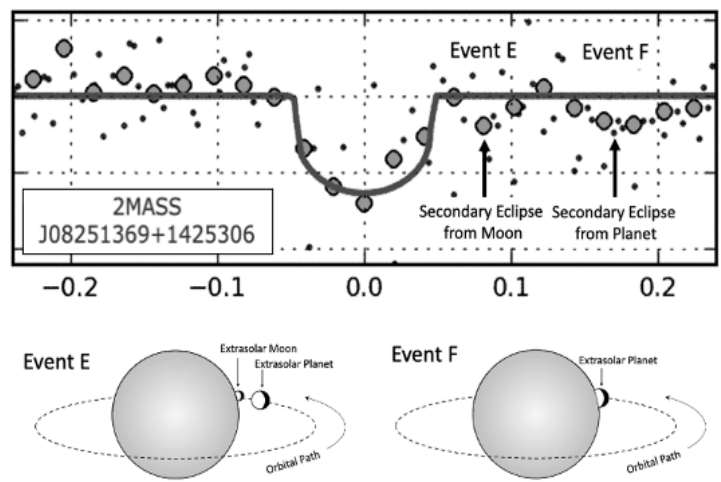

Figure 9: Post-transit light curve for first secondary eclipse (Event E) and second secondary eclipse (Event F).
(Source: EE Explorers and Planetary Sciences, Inc.)

## COMPETING HYPOTHESES
## FOR THE SOURCE OF THE PHOTON FLUX PERTURBATIONS

Numerous astrophysical processes occur during planetary transits that can account for the source of the duo-photon flux (double dip) associated with 2MASS J08251369+1425306. The Rossiter–McLaughlin effect, for illustration, is a natural phenomenon that occurs during a planetary transit and can generate dips in the light curve resembling the perturbation from the proposed extrasolar moon. The anomaly occurs in part due to the doppler reflex motion that an orbiting planet imparts on its rotating parent star (Figure 10a). As the parent star rotates on its axis, one side of its photosphere will be seen to be coming toward the viewer (blueshifted) while the other visible quadrant appears to be moving away (redshifted). When the extrasolar planet transits the parent star, it blocks part of the latter's disc, preventing some of the blue- and redshifted light from reaching the observer. As a result, the transit will appear as an anomalous radial-velocity variation on the light curve (Figure 10b).[32] Although some studies have speculated on the prospects of using K2 data to detect extrasolar planets using the Rossiter–McLaughlin effect, it will be difficult to detect transits using this method due to the limitations of Kepler's instruments.[33] Therefore, we ruled out the Rossiter–McLaughlin effect as the source of the perturbations.

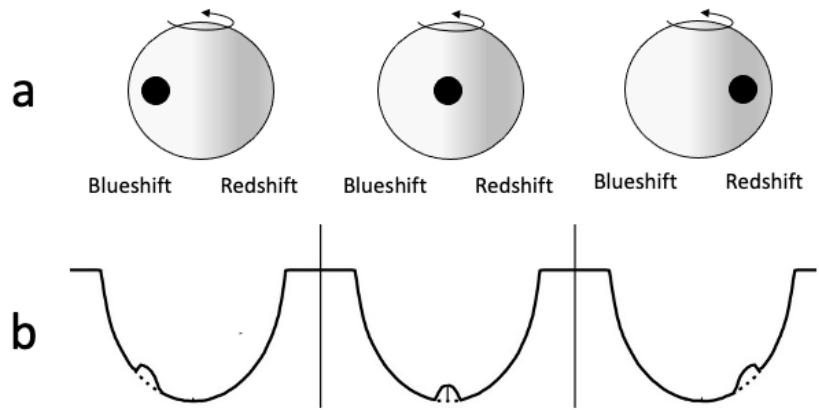

Figure 10: Simulation of Rossiter–McLaughlin effect (a) and
associated light curves (b). (Source: Planetary Sciences, Inc.)

Photons escaping the edge of the stellar disc, where the temperature is cooler due to altitude and the radiation is less intense, will cause the star to appear darker along its outer edges (Figure 11a).[34] This phenomenon, known as *limb darkening*, causes the photons emitted from the limb of the stellar disc to follow a more oblique path through the stellar atmosphere than the photons emitted from the center.[35] Limb darkening, however, depends on the spectrum (i.e., visible vs infrared) being used to observe the star.[36] The limb-darkening effect is largest at short wavelengths where a highly rounded light curve is observed. Conversely, as the effective stellar temperature increases, the transit light curve becomes shallower and wider (Figure 11b). However, for the purposes of this investigation, the effects of limb darkening have already been modeled into K2 light curves with limb darkening coefficients.

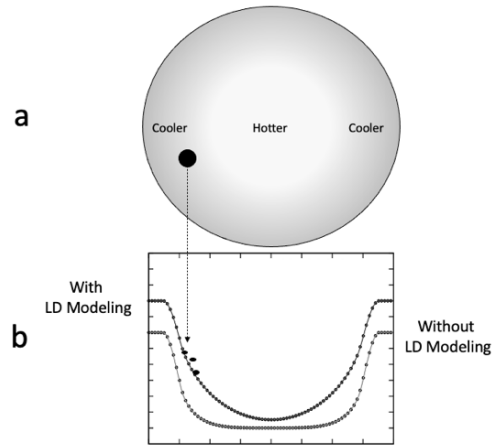

Figure 11: Simulation of limb-darkening effect on
light curves. (Source: Planetary Sciences, Inc.)

Our final hypothesis to account for the perturbations is sunspots—dark blotches on the surface of the star associated with intense magnetic activity. The temperatures of these sunspots are lower than the rest of the photosphere, which gives them their dark appearance.[37] As the star rotates, sunspots will come in and out of view causing changes in the star's brightness. The pattern in the star's light curve, therefore, will repeat once per rotation period of the star. As a result, the sunspot will produce a small bump in the transit light curve until the sunspot's lifecycle ends (Figure 12). However, because the light curve for 2MASS J08251369+1425306 was assembled from five separate transits, the sunspot should have formed multiple perturbations along the light curve—which was not evident in the K2 data we analyzed.

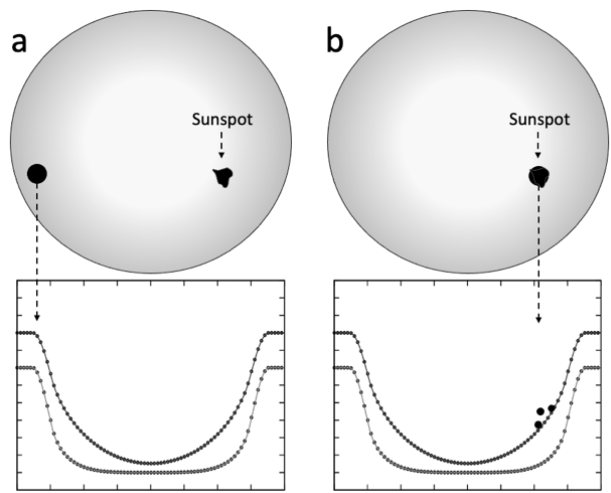

*Figure 12: A simulation of an extrasolar planet occulting a sunspot.*
(Source: Planetary Sciences, Inc.)

## CONCLUSIONS

Our investigation of K2 light curves established that extrasolar moons are detectable by inspecting the perturbations in the timing of transits produced by multiple celestial bodies orbiting their parent star. Of the $5 \times 10^3$ main-sequence stars we examined, at least eight of the stars revealed light curves analogous to the extrasolar moon-like perturbations demonstrated in our simulation. The two individual decreases in the photon flux associated with 2MASS J08251369+1425306, we postulate, imply the presence of an extrasolar planet with a radius of $R_P \approx 0.129$ and its companion, an extrasolar moon, with a radius of $R_M \approx 0.0048$. During our investigation, moreover, we analyzed four other astronomical anomalies that could account for the perturbations, namely the Rossiter–McLaughlin effect, limb darkening, a multiplanet system or the occultation of a sunspot on the surface of the star. A review of the main-sequence stars we examined, however, found no analogous K2 light curves to accommodate our competing hypotheses and/or account for the perturbations associated with 2MASS J08251369+1425306.

## BIOGRAPHY

Antonio Paris, the Principal Investigator for this study, is the Chief Scientist at Planetary Sciences, Inc., an Assistant Professor of Astronomy and Astrophysics at St. Petersburg College, FL, and a graduate of the NASA Mars Education Program at the Mars Space Flight Center, Arizona State University. He is a professional member of the Washington Academy of Sciences, the American Astronomical Society and The Explorers Club; and his latest peer-reviewed publication focused on broad-line quasars—active galactic nuclei emissions powered by supermassive black holes.